\documentclass[a4paper,notitlepage,twocolumn]{biophys_customised}
\usepackage{dcolumn,color,helvet,times}
\usepackage{amsmath,lastpage,bm,textcomp}
\gridframe{N}
\cropmark{N}
        
\usepackage{graphicx}
\newcommand{\abs}[1]{\ensuremath{\left\vert#1\right\vert}}
\newcommand{\ket}[1]{\ensuremath{\vert#1\rangle}}
\newcommand{\bra}[1]{\ensuremath{\langle #1\vert}}

\newcommand{\kb}[2]{\ensuremath{\vert #1 \rangle \langle #2 \vert}}

\newcommand{\rad}{\raisebox{0.6ex}{\ensuremath{\bullet}}}

\def\beq{\begin{equation}}
\def\eeq{\end{equation}}
\def\beqa{\begin{eqnarray}}
\def\eeqa{\end{eqnarray}}

\begin{document}

\vspace{2cm}
\title{A new type of radical-pair-based model for magnetoreception}

\author{A. Marshall Stoneham,*$^{,}$\authdagger$^{,1}$  Erik M. Gauger,*$^{,2,3}$ Kyriakos Porfyrakis,$^{,2}$ Simon C Benjamin,$^{,2,3}$ \\ and Brendon W Lovett$^{,4,2}$}

\address{$^{1}${\it Department of Physics and Astronomy, University College London, Gower Street, London WC1E 6BT, United Kingdom}}
\address{$^{2}${\it Department of Materials, University of Oxford, Oxford OX1 3PH, United Kingdom}}
\address{$^{3}${\it Centre for Quantum Technologies, National University of Singapore, Singapore}}
\address{$^{4}${\it School of Engineering and Physical Sciences, Heriot Watt University, Edinburgh EH14 4AS, United Kingdom}}
\address{*These authors contributed equally to the work presented here.}
\address{{\addrdagger}Deceased on 18th February 2011.}

\maketitle{}

\vspace{1cm}

\noindent ABSTRACT Certain migratory birds can sense the Earth's magnetic field. The nature of this process is not yet properly understood. Here we offer a simple explanation according to which birds literally `see' the local magnetic field through the impact of a physical rather than a chemical signature of the radical pair: a transient, long-lived electric dipole moment. Based on this premise, our new picture can explain recent surprising experimental data indicating long lifetimes for the radical pair. Moreover, there is a clear evolutionary path toward this field sensing mechanism: it is an enhancement of a weak effect that may be present in many species.

\clearpage
\twocolumn

\section{Introduction}

It is well established that certain migratory birds can detect the direction of the Earth's magnetic field, and use this as a compass for orientation \cite{mouritsen05,johnsen08,ritz10}. An obvious explanation for this remarkable ability would be the use of magnetised materials in the bird's body, which is likely to change orientation with the external field \cite{fleissner03, falkenberg10}. However for species such as the European Robin the evidence points to a very different mechanism: the prevailing hypothesis is that field orientation is initially detected through its influence on photo-excited electronic spins. Photons are evidently important since the birds can only orientate in a magnetic field when light (which may be dim) is available \cite{wiltschko72, wiltschko02, ritz04}, and with an undamaged visual system \cite{zapka09}. A neuronal pathway that is likely responsible for the processing of light-dependent magnetic information was suggested by Ref.~\cite{heyers07}. Meanwhile a recent observation provides strong support for the role of electron spins: the birds' are disorientated by a weak oscillatory field whose frequency is close to the resonant frequency for an electron in the Earth's magnetic field \cite{ritz04, ritz09}. While these detailed results come from migratory bird studies, there is evidence of a similar sensitivity in non-migratory birds, such as chickens~\cite{wiltschko07} and zebra finches~\cite{keary09}, and in other animal phyla, as evidenced by experiments on the American cockroach~\cite{vacha09}. This suggests that this form of magnetoreception may occur in diverse organisms.

These findings have led to the popularity of the Radical Pair (RP) model \cite{ritz04, ritz00, maeda08, rodgers09a, cai10}, which begins to explain how light activated magnetic sensing could happen. The central feature of the model is an optical excitation of certain biomolecules which leaves a fraction of these molecules in a spin triplet state, with a spatially-separated pair of spins. In magnetically anisotropic systems, the number of spin triplets depends on the orientation of the magnetic field. If the mole\-cules are themselves (at least partly) oriented~\cite{hill09,lau09,solovyov10} and if the bird can somehow detect the relative population of spin triplets, then an optically activated avian compass is possible. 

However, the transduction mechanism by which electron spin states translate to a macroscopic signal is not well understood. A typical explanation is that some signature chemical is synthesised only when the triplet state decays. This chemical may then interfere with the normal process of vision, or it might be detected by some independent sensor structure in the eye \cite{ritz10}. 
Explanations of this kind are puzzling for two reasons: 

First, they involve a complex chemistry which must have evolved within the eye, independent of (but consistent with) the process of normal vision. Yet, no sensory additional magnetoreception receptors have yet been identified. 

Second, this model would seem to function best when the cycle time, i.e. the time for production of the signature chemical(s) or photons, is short -- shorter cycles would lead to higher rate of production and thus better signal/noise ratios. However, in the real system it seems that the opposite is true: the RP lifetime, as measured by spin resonance experiments on live birds, is extraordinarily long \cite{ritz09, liedvogel07, gauger11}.

At least one simpler alternative to the chemical transduction mechanism has been proposed~\cite{leask77}, but this did not explain the then-unrecognised need for long triplet lifetimes. Here, we will describe a model of the compass in which it is straightforward to understand that need; indeed this property is so crucial that the molecules involved could have evolved through natural selection of slow electronic decay rates. We will further describe why no apparatus for detecting chemical products is required in our model. In essence, the Earth's magnetic field translates to a local electrostatic (or strain) field which directly modulates vision. This may be seen as an evolutionary enhancement of an inherent sensitivity, analogous to the well-studied `Haidinger's brush' \cite{shute74} phenomenon, in which the polarization of light (as opposed to the direction of a magnetic field) is detected through a molecular electric dipole transduction mechanism. To the best of our knowledge, direct evidence for local electrical fields affecting the visual process has not (yet) been reported, however, in vitro laboratory studies on pertinent biomolecules show that such effects are possible in principle under realistic circumstances as we presently discuss. In addition, dedicated sensory systems for detecting electric fields do exist in certain species, e.g.~sharks and rays are known to be able to detect extremely weak external electric fields as low as $1~\mu\mathrm{V/m}$ \cite{kalmijn71}. 

We believe that our  version of the RP model is a simpler and more complete hypothesis than previously proposed models. Importantly, our model is equally consistent with all experimental findings whilst also providing a sound evolutionary pathway to the observed long RP lifetime.

\section{Model}

We start with a general description of our compass model before turning to a specific set of example parameters to demonstrate the feasibility of our proposed mechanism. Our model features several of the successful tenets of the conventional RP model. First, charge-separated radical pairs are created when light is absorbed by the compass molecule. Second, the radical pair formed within or from the compass molecule is magnetically anisotropic and (at least partially) aligned, and thus the relative population of RPs ending up in triplet and singlet configurations depends on the orientation of the molecule to Earth's magnetic field. Third, the singlet RP can decay directly back to the ground state, but for the triplet this route is blocked. However, our model differs in the important aspect of the compass signal transduction. In contrast to previous proposals for RP based magnetoreception, here the signal is not of a chemical nature, but is rather a physical effect that is associated with the decay of the spin triplet state to a long lived charge separated state, which serves as the bird's signalling state. More specifically, the normal vision system \cite{lewis88} is modulated by the electric field of the electric dipole which accompanies this charge-separated triplet signalling state. This requires the compass molecules to be located directly on the retina. However, the retina also seems the most likely location in the standard RP model due to the fact that it is already integrated with a system to initiate signals to the brain in normal vision.

\begin{figure}
\begin{center}
\includegraphics[width=\columnwidth]{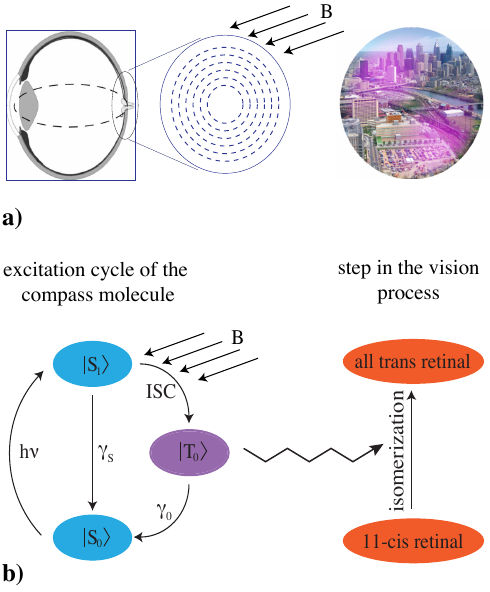}\\
\caption{\textbf{a:} Left: Schematic diagram showing how the molecules might be aligned in the retina; a combination of light and a magnetic field could induce dipole moments for certain molecular orientations. Right: These dipoles would create an electric field which would allow the bird literally to `see' the magnetic field direction.
\textbf{b:} Cycle of the compass molecule: 
following photoexcitation from \ket{S_{0}} to \ket{S_{1}}, the branching ratio of direct relaxation into the ground state or via a long-lived triplet state \ket{T_{0}} depends on the orientation of the molecule with the geomagnetic field. Here, \ket{T_{0}} is a charge separated state with an electric dipole moment, thus affecting the isomerization of retinal, which is a crucial step of the visual process.}
\label{fig:Cartoon}
\end{center}
\end{figure}

We base our model on a compass molecule with the following properties. First, it possesses an (optically) excited singlet state that can evolve into a triplet state, dependent on an external magnetic field. This phenomenon is well-established in artificial systems, such as for example, self-trapped excitons in alkali halides \cite{itoh00} and NV$^{-}$ centres in diamond \cite{lai09}. Second, we require a lower lying meta-stable charge-separated triplet state with a sufficient lifetime to influence the visual process in the retina. Sufficiently long triplet lifetimes of a few milliseconds are not uncommon in photo-active molecules \cite{liedvogel07}. Third, we require a spin level structure that allows this long-lived triplet state to be dephased by a resonant RF field. We shall discuss one possible mechanism for this later. Fourth, as in the standard RP model, the molecules should form an ordered structure on the surface of the retina, although some amount of disorder can be tolerated \cite{hill09,lau09,solovyov10}. In Fig.~\ref{fig:Cartoon} we display a circular arrangement, but we note the actual pattern in the bird's eye could be different and our mechanism does not rely on any particular pattern. However, note that in humans a circular arrangement similar to the one shown in Fig.~\ref{fig:Cartoon} has been proposed for lutein molecules as a possible explanation of the fact that some people can directly see light polarization \cite{shute74, bone83}. In this case it has been speculated that the alignment could originate from the known radial orientation of nerve fibres. As will become clear shortly, an elongated shape of the radical pair, e.g. brought about by a rod-like compass molecule, will be desirable to obtain a large radical pair electric dipole.

The magnetic orientation sensing mechanism then proceeds as follows: optical excitation gives rise to the formation of the charge-separated triplets - the signalling states - for certain orientations of the magnetic field with respect to the molecular axes. The result is an electric field distribution on the retina which reflects the orientation of the magnetic field. More experimental data is available from studies of bacteriorhodopsins than animal rhodopsins, but the structure of the two are similar despite probably having evolved independently. In particular, both classes have an identical light absorbing chromophore, the 11-cis-retinal, whose photoisomerization is the primary event in their photochemical cycles. Relatively weak electric fields between $10^{5} -10^{7}~\mathrm{V/m}$ affect the photoenergetic reaction and absorption spectrum of bacteriorhodopsin \cite{birge90,borisevitch79, lukashev80}, as well as the \textit{cis}-to-\textit{trans} isomerization of many other complex molecules \cite{wahadoszamen04, tong08}. In Ref.~\cite{schenkl05} a link between electric field generation and isomerization of retinal in bacteriorhodopsin was etablished. It therefore seems plausible, even likely, that there will be an electric effect on retinal isomerization in avian rhodopsin.

An electron-hole dipole with average charge separation of only one nanometre produces an electric dipole field with magnitude $10^{6}~\mathrm{V/m}$ up to a distance of $10~\mathrm{nm}$, while a field of the order of $10^{5}~\mathrm{V/m}$ even extends to $25~\mathrm{nm}$. Each compass dipole thus possesses a sizeable `sphere of influence' in which it could directly affect the photoisomerization of retinal~\cite{kobayashi01}, meaning the bird would literally be able to `see' the magnetic field as a superimposed feature in its normal visual image.

\begin{figure}
\begin{center}
\includegraphics[width=\columnwidth]{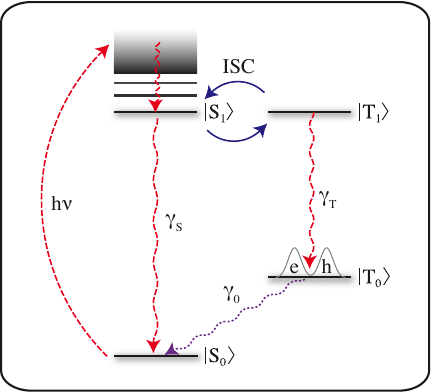}
\caption{Schematic diagram showing the key features of the compass mechanism: the system relaxes back into the singlet ground state \ket{S_{0}} after most photoexcitation events. However, there is a small intersystem crossing rate, which depends on the orientation of an asymmetric $g$-tensor with the geomagnetic field. Population in \ket{T_{1}} relaxes into a long-lived triplet state \ket{T_{0}}.
Transitions between different charge states are depicted by dashed arrows, while a change of the spin state is shown by a solid arrow. The (dotted) decay from \ket{T_{0}} to \ket{S_{0}} involves both a charge and a spin transition.}
\label{fig:Schematic}
\end{center}
\end{figure}

Let us now consider an example system that would exhibit the features required by our model. We will describe the simplest possible molecular energy level structure required for our proposed mechanism, though it is of course likely that any real system will have extra features. The scheme we have in mind possesses four relevant energy levels as sketched in Fig.~\ref{fig:Schematic}.

We imagine that our biomolecule, like most others, has a singlet ground state \ket{S_{0}}. Light can excite population to higher lying singlet states, as these transitions are strongly allowed by dipole selection rules. The molecule may then experience a cascade of non-radiative decays followed by a charge transfer during which the electron and hole become spatially separated, forming the radical pair singlet state \ket{S_{1}}. 
The singlet \ket{S_{1}}  would normally simply decay back to the ground state (after a time that may be as short as a nanosecond), but it is also possible that population branches off into a (degenerate or close to degenerate) a radical pair triplet state \ket{T_{1}}. As we discuss below, the rate of the singlet-triplet interconversion -- or intersystem crossing (ISC) rate -- can be dependent on the geomagnetic field, and lies at the heart of the magnetoreception mechanism. 
Finally, \ket{T_{0}} is a lower-lying, long-lived triplet state that is reached through a fast optical or non-radiative decay of \ket{T_{1}}. The charge-separation of electron and hole in the radical pair configuration means that population in the level  \ket{T_{0}} has an associated electric dipole moment, and this triggers the visual stimulus for the compass. [We note that our model requires a charge separation only for the level \ket{T_{0}}.  It seems likely that the formation of the (charge-separated) radical pair occurs in the relaxation to \ket{S_{1}} and persists until the ground state \ket{S_{0}} is reached. However, in principle, our model would also allow for the levels \ket{S_{1}} and \ket{T_{1}} to have a localised excitonic character, with the spatial separation of charges into the radical pair only happing in the relaxation from \ket{T_{1}} to \ket{T_{0}}.]
 
The dependence of the intersystem crossing rate on the geomagnetic field must be associated with an anisotropic term in the Hamiltonian. The origin of this term is not important; a hyperfine coupling between the electron in the optically excited exciton and a nuclear spin has been widely proposed in the literature \cite{ritz00, rodgers09a}. In order to keep our discussion as simple as possible, we will assume that \ket{S_1} and \ket{T_{1}} are subject to an isotropic electron $g$-factor $g_{e} = 2$ and  a uniaxially anisotropic hole tensor: 
\begin{equation}
\mathbf{g}_{h} = \left( \begin{array}{ccc}
 2 + \delta g & 0 & 0 \\
 0 & 2 + \delta g & 0 \\
 0 & 0 & 2 + \Delta g
\end{array}
 \right),
\end{equation}
and as specific examples, we assume $\delta g = 0$ and $\Delta g = 0.2$. Here we have used the language of excitons, which are conventionally pictured as consisting of a single excited electron and a missing ground state electron (or hole); in a radical pair picture we would equivalently say that the two $g$-factors apply to the two unpaired radical spins. We have checked that the qualitative predictions of our model also work for the case of an anisotropic hyperfine coupling similar to the one  described in Ref.~\cite{gauger11}. 

Electric dipole selection rules mean that, following the photoexcitation, the system is found in the pure singlet state \ket{S_{1}} which is degenerate with the triplet level \ket{T_{1}}. We write for the Hamiltonian at this stage of the process:
\begin{equation}
H_{ISC} = \frac{1}{2} \mu_{B} \left( g_{e} \mathbf{B} \cdot \mathbf{S}_{1} + \mathbf{B} \cdot  \mathbf{g}_{h} \cdot \mathbf{S}_{2} \right),
\label{eq:ISCHamiltonian}
\end{equation}
where $ \mu_{B}$ is the Bohr magneton, $\mathbf{B}$ is the magnetic field vector, and $\mathbf{S}_{i} = (\sigma_{x}, \sigma_{y}, \sigma_{z})_{i}$ is the spin operator for electron ($i=1$) and hole ($i=2$). The factor $1/2$ accounts for the fact that all our Pauli matrices have eigenvalues $\pm 1$. The magnetic field strength in Frankfurt (the site where the relevant experiments were performed~\cite{ritz04, ritz09}) is  $B_{0 } = 47~\mu\mathrm{T}$. The field's orientation with respect to the $g$-tensor is determined by the angles $\theta$ and $\phi$, \mbox{$\mathbf{B} =  B_0 (\cos \phi \sin \theta, \sin \phi \sin \theta, \cos \theta)$}.

Based on Hamiltonian (\ref{eq:ISCHamiltonian}), we obtain the following matrix elements for the three triplet sublevels $\ket{t_{+}} = \ket{\uparrow\uparrow}$, $\ket{t_{0}} = (\ket{\uparrow\downarrow} + \ket{\downarrow\uparrow})/\sqrt{2}$ and $\ket{t_{-}} = \ket{\downarrow\downarrow}$ of \ket{T_{1}}:
\begin{eqnarray}
\bra{S_{1}} H_{ISC} \ket{t_{0}} & = & \mu_{B} B_{0} \, \Delta g \, \cos \theta \label{eq:DeltaElement}, \\ 
\bra{S_{1}} H_{ISC} \ket{t_{\pm}} & = & \pm \mu_{B} B_{0} \, \delta g \,  \sin \theta e^{\mp i \phi} / \sqrt{2}.
\end{eqnarray}
To obtain a signal that depends on the relative orientation of $g$-tensor and field, as is required for a compass, we must thus have $\delta g \neq \Delta g$, a condition which is fulfilled by our particular choice of parameters. Owing to the axial symmetry of Hamiltonian (\ref{eq:ISCHamiltonian}), $\phi$ is unimportant and we need only consider $\theta \in [0, \pi/2]$.

We now use a phenomenological Lindblad master equation~\cite{breuer02} to model the evolution of the density matrix that describes the quantum dynamics of our (open) molecular system. The optical excitation between \ket{S_{0}} and \ket{S_{1}} is modelled as an incoherent process with a Lindblad operator $P_{X} = \kb{S_{1}}{S_{0}}$ with associated rate $\gamma_{X}$. Similarly, the decay events are described by Lindblad operators $P_{S} = \kb{S_{0}}{S_{1}}, P_{T} = \kb{T_{0}}{T_{1}}$ and $P_{0} = \kb{S_{0}}{T_{0}}$ with respective rates $\gamma_{S}, \gamma_{T}$ and $\gamma_{0}$  as depicted in Fig.~\ref{fig:Schematic}. Using only the matrix element Eq. (\ref{eq:DeltaElement}) as the effective Hamiltonian $H$ and all of the above Lindblad operators, we obtain as the master equation governing the time evolution of the  system's density matrix $\rho(t)$~\cite{nielsen00, breuer02}:
\begin{eqnarray}
\dot{\rho} =  -\frac{i}{\hbar} [H, \rho ]  +  \sum_{i} \gamma_{i} \left( P_i \rho P_i^{\dagger}   - \frac{1}{2} \left( P_i^{\dagger} P_i \rho + \rho  P_i^{\dagger} P_{i} \right) \right).
\label{masterEqn}
\end{eqnarray}
We are interested in the steady state population $\mathcal{T}$ of the charge-separated triplet level \ket{T_{0}}, which is found by setting the LHS of Eq.~(\ref{masterEqn}) to zero, yielding
\begin{equation}
\mathcal{T} = \frac{4 \gamma_{X} g(\theta)}{ 4 \gamma_{X} g(\theta) + \gamma_{0} \Gamma \left( (\gamma_{S} + \gamma_{X}) \hbar^{2} + \frac{4}{\gamma_{T}} \left( 1+ \frac{2 \gamma_{X}}{\Gamma} \right) g(\theta) \right)}
\end{equation}
where $\Gamma = (\gamma_{S} + \gamma_{T})$ and $g(\theta) =  \abs{\bra{S_{1}} H_{ISC} \ket{t_{0}} }^{2} $. In the regime of interest, the lifetime of \ket{T_{0}} is much longer than that of the excited states \ket{S_1} and \ket{T_1}, i.e. $\gamma_{0} \ll \gamma_{S} \approx \gamma_{T}$.
It is also reasonable to assume that $\gamma_{X} \ll \gamma_{S}$ and $\gamma_{S}  \gg   \mu_{B} B_{0} \Delta g / \hbar$, and to a good approximation we therefore find
$\mathcal{T} \propto g(\theta)/\gamma_0$. Importantly, the steady population $\mathcal{T}$ is thus largely independent of specific values for any of the decay rates except $\gamma_0$ as long as the hierarchy assumed above is fulfilled.

\section{Results}

\begin{figure}
\begin{center}
\includegraphics[width=\columnwidth]{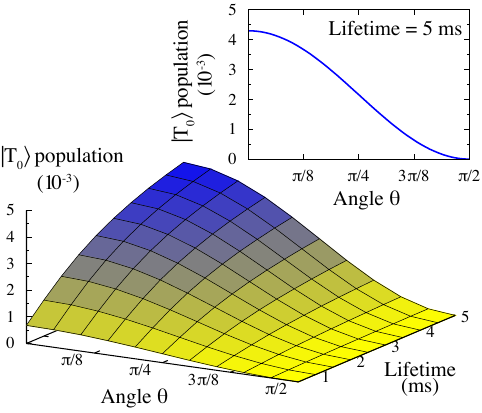}
\caption{Steady state population of \ket{T_{0}}, $\mathcal{T}$, is shown as a function of this state's lifetime $1/\gamma_{0}$ and the angle $\theta$ between the axial $g$-tensor and the Earth's magnetic field for $\Delta g = 0.2$. For other parameters see main text; note that $1/\gamma_{0}$ is assumed to be long in contrast to the lifetimes of \ket{S1} and \ket{T1} which may be as short as $1~\rm{ns}$. The 2D-plot in the upper right follows the $\cos^{2} \theta$ dependence of the squared relevant matrix element [see Eq.~(\ref{eq:DeltaElement})].}
\label{fig:TripletPop}
\end{center}
\end{figure}

Fig.~\ref{fig:TripletPop} shows a full numerical solution of the triplet population $\mathcal{T}$ as a function of the angle $\theta$ and the triplet lifetime $1/\gamma_{0}$. Here, we have assumed the lifetimes of the \ket{S_{1}} and \ket{T_{1}} states are $1~\mathrm{ns}$, and used an excitation rate $\gamma_{X} = 10^{6}~\mathrm{s}^{-1}$. As shown in Fig.~\ref{fig:TripletPop} a longer excited lifetime would increase the average electric field (i.e.~the product of the molecule's dipole moment and the number of dipoles present at any time), but importantly, it also gives each individual dipole more time to have an effect on other processes in the vision system such as the isomerization of any nearby retinal. We would therefore expect that birds whose signalling states persist longer --  by means of a longer spin coherence time -- would be able to see the magnetic field with more contrast; an evolutionary advantage that could have occurred in small increments through natural selection. Note that for our model to work, we must assume that both spins in the \ket{T_0} state are devoid of hyperfine-, exchange- and dipole-coupling-induced spin flip-flops on a sub-millisecond time\-scale, as these could enable a faster relaxation back to the \ket{S_0} state by corrupting the triplet spin state (see also the discussion in the following two paragraphs).

Recent experiments show that a very small oscillating magnetic field can disrupt the bird's ability to orient~\cite{ritz04, ritz09}. In Ref.~\cite{ritz09}, the authors report that a perturbing magnetic field of frequency of 1.316~MHz (i.e.~the resonance frequency of an electron spin for a $g$-factor of 2) with a field strength of only $15~{\rm nT}$ suffices to completely disorient the birds. The MHz frequency immediately implies a bound on the time of the process (since the field would appear static for a sufficiently rapid decay back to the molecule's ground state in less than a microsecond). Moreover, considering the oscillating magnetic field strength implies a much longer process time of at least several hundred microseconds to give the weak radiofrequency field sufficient time to affect the spin state~\cite{gauger11}. Importantly, as discussed above our proposed transduction mechanism provides a motivation for such a long process time, including the need for faithfully preserving the triplet spin state in order to block premature relaxation from the signalling state \ket{T_0} to the ground state \ket{S_0}.

\begin{figure}
\begin{center}
\includegraphics[width=\columnwidth]{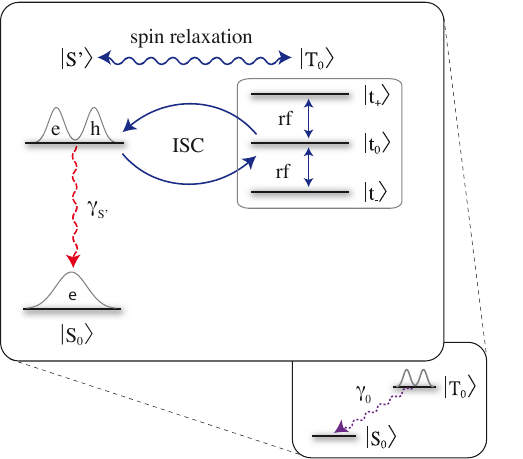}
\caption{Possible explanation of the relaxation from \ket{T_{0}} to the ground state \ket{S_{0}}:  the charge relaxation follows a spin transition to the auxiliary singlet level \ket{S'}. Population in \ket{t_{0}} is connected to \ket{S'} via an ISC and thus decays quickly, whereas population in \ket{t_{\pm}} is trapped for the duration of the spin coherence time. However,  a resonant RF field mixes the spin states, leading to faster relaxation of the entire triplet population. }
\label{fig:TripletMixing}
\end{center}
\end{figure}

There are numerous explanations by which a weak RF field could plausibly disrupt the compass mechanism by shortening the lifetime of such a long-lived triplet state. Essentially, whenever the RF field only rotates one of the two spins of the triplet because the two spin transitions are not degenerate due to different g-factors or environmental couplings, a fast decay route to the ground state becomes available by  converting the triplet into a singlet radical pair. In the following, we discuss one simple possibility. We focus on the lower-lying triplet to ground state transition. For understanding the directional sensitivity we needed no more than those two levels, but in order to understand the resonant effects of a magnetic field we must now explicitly include the distinct spin states, see Fig~\ref{fig:TripletMixing}. We distinguish the specific triplet states $\ket{t_+}$, $\ket{t_0}$, $\ket{t_-}$, and we label the corresponding singlet level as $\ket{S'}$. This latter level will have a fast, spin-allowed decay to the molecule's ground state $\ket{S_0}$. State \ket{S'} thus separates the spin and the charge transition of this process. Importantly, the auxiliary level could be eliminated from the dynamics so long as the decay $\gamma_{S'}$ is large enough, reducing the model once more to the simpler picture displayed in Fig.~\ref{fig:Schematic}.

In our illustrative example, we assume an electron $g$-factor of $g_{e} = 2$ and $g_{h} = 2.2$ for the hole (both isotropic). Working in the basis where the $z$-axis is defined by the applied static magnetic field direction, the triplet now has the three sublevels shown in Fig.~\ref{fig:TripletMixing}, and the Hamiltonian in an RF field reads:
\begin{eqnarray}
H_{\rm RF} = \frac{1}{2} \mu_{B} \big[  B_{0} & \left( g_{e} \sigma_{z1} +  g_{h}  \sigma_{z2} \right) & \nonumber \\
 +   B_{\rm RF}   \cos \omega t & \left( g_{e} \sigma_{x1} +  g_{h}  \sigma_{x2} \right) & \big].
\label{eq:RFHamiltonian}
\end{eqnarray}
Here, the oscillatory field of strength $B_{\rm RF} = 150~\mathrm{nT}$ and frequency $\omega$ is applied orthogonal to the static field $B_{0}$, since only its perpendicular component affects the compass~\cite{ritz09}.We emphasize that the magnitude of $B_{\rm RF}$ is ten times larger than the smallest value that has been reported to disrupt the avian compass. If our model is valid then for this stronger RF field we must certainly find that our predicted orientational effect is washed away. Note that the three sublevels \ket{t_{+}}, \ket{t_{0}} and \ket{t_{-}} of Fig.~\ref{fig:TripletMixing} take a different form to the sublevels of \ket{T_{1}} that we have considered before, whose axis was defined by the anisotropic $g$-tensor of the hole. (Following a spin-preserving relaxation from \ket{T_{1}} to \ket{T_{0}}, we then obtain the state $\ket{T_{0}} = \cos \theta \ket{t_{0}} + \sin \theta (\ket{t_{+}} - \ket{t_{-}}) \sqrt{2}$ written in terms of the triplet sublevels.) 

\begin{figure}
\begin{center}
\includegraphics[width=\columnwidth]{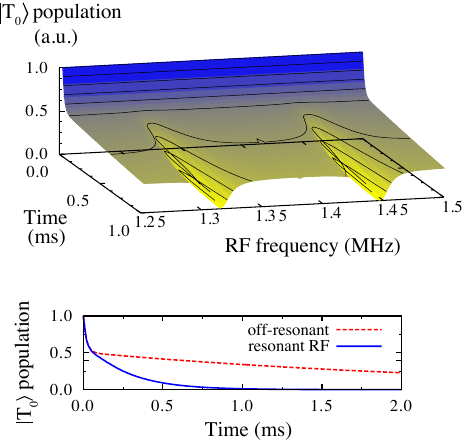}
\caption{Surviving \ket{T_{0}} population as a function of time and RF frequency $\omega$. After an initial fast decay of the \ket{t_{0}} population (see text), a slowly decaying plateau of triplet population is reached. However, on resonance with the either the electron or the hole spin, the oscillatory field  drastically shortens the triplet lifetime. See text for parameters. }
\label{fig:RFDecay}
\end{center}
\end{figure}

Fig.~\ref{fig:RFDecay} shows the surviving \ket{T_{0}} population as a function of time. This was obtained by using Hamiltonian (\ref{eq:RFHamiltonian}) in a general Lindblad master equation [see Eq.~(\ref{masterEqn})] with appropriate Lindblad operators. In this case, the operators are $P_{S'} = \kb{S_{0}}{S'}$ for the decay, and $P_{x,i} = (\sigma_{x})_{i}$ for describing spin flip decoherence  of electron and hole spin, respectively. The decay rate is $\gamma_{S'} = 10^{5}~\mathrm{s}^{-1}$ and we assume slower but equal spin decoherence rates of $0.2 \times 10^{3}~\mathrm{s}^{-1}$. As the initial state, we assign half of the population to \ket{t_{0}} and the other half is equally distributed between the sublevels \ket{t_{+}} and \ket{t_{-}}, consistent with the state resulting from a $\ket{T_1}$ decay. 

The pronounced kink at short times in the data of  Fig.~\ref{fig:RFDecay} shows how the \ket{T_{0}} relaxation proceeds in two stages: population in \ket{t_{0}} undergoes a direct ISC to \ket{S'}, subsequently decaying to the ground state in much less than a millisecond. The other half of the triplet population in the \ket{t_{\pm}} sublevels survives much longer, until all spin states are eventually mixed by the $P_{x,i} = (\sigma_{x})_{i}$ spin decoherence processes. Only when the oscillatory field is resonant with either the electron or hole spin (at 1.316 or 1.447 MHz  due to the different $g$-factors), are all triplet levels mixed on a sub-millisecond timescale, leading to a fast decay of the \textit{entire} triplet population. The corresponding two valleys in the data elucidate how the resonant RF field thus severely reduces the excited triplet lifetime. Such a reduction of the \ket{T_{0}} lifetime would plausibly affect the compass mechanism and disorient the bird.

We note that the general behaviour displayed in Fig.~\ref{fig:RFDecay} does not depend on the specific choice of parameters, so long as the hierarchy of the various processes is preserved. Indeed, one can also think of entirely different physical mechanisms for the \ket{T_{0}} relaxation that would be equally consistent with experimental observations. Crucially, all possible explanations depend on an excited state with a lifetime of more than a millisecond, as the oscillatory field strength is simply too weak to significantly affect the spin state in a shorter duration.

\section{Discussion}

A widely proposed molecule for the RP mechanism in birds is cryptochrome, though there is currently only indirect evidence that this is indeed the molecule responsible for magnetoreception~\cite{rodgers09a, henbest08}. Further, a recent study suggests that cryptochrome is arranged on the retina in close proximity to the UV cones in a fashion fulfilling the requirements of RP based magnetoreception \cite{Niessner11}.
Despite the current uncertainty about the role of cryptochromes~\cite{liedvogel10}, we feel a presentation of how our model might be realized in cryptochromes would be helpful. These molecules consist of a flavin adenine dinucleotide (FAD) and a light-harvesting cofactor. The ground state \ket{S_0} is thought, in the avian retina, to correspond to the FAD in its fully oxidized state. Optical excitation then induces electron transfer from a chain of three tryptophan residues to the FAD, which causes reduction to the radical FADH\rad. This state would now correspond to \ket{S_1}, and can undergo ISC through anisotropic hyperfine interactions. Straightforward dipole-allowed recombination to the ground state is possible for the singlet, but blocked for the triplet. The signalling state is the only one open to the triplet,~\cite{rodgers09a} and in our case we would need a long-lived charge separated state. This could possibly be formed through charge transfer to a tyrosine residue~\cite{aubert99}, though whether the electric dipole so formed would survive for long enough for our purposes before (de)-protonation would need further study.

The model we have proposed could be tested, and we now discuss some possible experiments that might be performed. First, it would be very interesting to probe the RF disruption mechanism further. In particular, if one could test the ability of a bird to navigate in an RF field across a range of static fields, it should be possible to obtain the width of the triplet resonance line that causes the disruption. Indeed, we would expect to find (at least) two resonances, one corresponding to the electron resonance and the other to the hole resonance (in a radical pair picture, the hole resonance would be the resonance of the second unpaired radical spin). A full mapping of the occurring resonances and their frequency dependence on changes in the external magnetic field would help to answer the question whether the directionality of the compass molecule is provided by an anisotropic g-factor or by an anisotropic hyperfine tensor. A second possibility is that birds using the mechanism we propose would be very sensitive to the polarization of the light used to induce the magnetoreception. The rod like molecules discussed in this paper would be expected to have a highly anisotropic electric susceptibility, and so the method we propose might only work in polarized light for molecules of certain orientations. It may be that the birds can adapt to this, but by changing the polarization of the light to which the birds are exposed periodically any adaption could be prevented, and again we might expect to see disruption of magnetoreception. In vitro experiments showing an electroluminescence would help to differentiate our proposed method from the previous RP based models. Precursor experiments similar to the artificial compass demonstrated by Maeda \textit{et al.} \cite{maeda08} could first be performed on candidate molecules in the laboratory to identify the specific sets of parameters for experiments on live birds.

We have proposed a comprehensive model that would allow a bird to sense the direction of the Earth's magnetic field. It relies only on processes that are common to standard ideas of vision mechanism and processes that exploit perfectly standard features of small biomolecules; nothing exotic is required. Further, we have suggested that a long lived triplet, essential for understanding the observed disruption of the effect by very weak radio waves, could have evolved through natural selection. Interestingly a number of human biomedical disorders are attributed -- with many doubts and reservations -- to low intensity, oscillating, electromagnetic fields in the same frequency range that disrupts magnetism-based bird navigation~\cite{adair00, robertson09}.

\section{Acknowledgements}
We thank Dr L. B. Leaver, Prof P. J. Hore and Prof~W.~B.~Motherwell for useful discussions. We acknowledge financial support from the National Research Foundation and Ministry of Education, Singapore, EPSRC UK, the QIPIRC (no. GR/S\-82176/01), and the Royal Society. This material is based upon work supported by DARPA under Award No.N66001-11-1-4100. Any opinions, findings, and conclusions or recommendations expressed in this publication are those of the author(s) and do not necessarily reflect the views of DARPA.

\end{document}